\documentclass[reprint,floatfix,amsmath,amssymb,superscriptaddress]{revtex4-1}
\usepackage{natbib}
\usepackage{physics}
\usepackage{graphicx}
\usepackage{bm}
\usepackage{epstopdf}
\usepackage{xr}
\usepackage{float}
\usepackage{lineno}
\usepackage{siunitx}
\DeclareSIUnit\Oersted{Oe}
\DeclareSIUnit\electronvolt{eV}
\DeclareSIUnit\emu{emu}
\usepackage{xcolor}
\usepackage[normalem]{ulem}
\usepackage{upgreek}
\usepackage{comment}

\newcommand{\kaiser}{cm$^{-1}$}

\begin{document}

\title{Helicity resolved Raman spectroscopy of mono- and a few-layers thick PtSe$_2$}


\author{Isamu Yasuda}
\affiliation{Department of Physics, The University of Tokyo, Bunkyo-ku, Tokyo 113-0033, Japan}

\author{Takuya Kawada}
\affiliation{Department of Physics, The University of Tokyo, Bunkyo-ku, Tokyo 113-0033, Japan}

\author{Hiroki Matsumoto}
\affiliation{Department of Physics, The University of Tokyo, Bunkyo-ku, Tokyo 113-0033, Japan}

\author{Masashi Kawaguchi}
\affiliation{Department of Physics, The University of Tokyo, Bunkyo-ku, Tokyo 113-0033, Japan}

\author{Masamitsu Hayashi}
\affiliation{Department of Physics, The University of Tokyo, Bunkyo-ku, Tokyo 113-0033, Japan}
\affiliation{Trans-scale quantum science institute (TSQS), The University of Tokyo, Bunkyo-ku, Tokyo 113-0033, Japan}

\date{\today}

\begin{abstract}
We studied helicity resolved Raman scattering in PtSe$_2$ flakes with different thicknesses.
The peak amplitude of helicity-switched Raman scattering is significantly larger than that of helicity-conserved scattering for the in-plane $E_{g}$ mode, consistent with the Raman tensor analyses and conservation law of angular momentum.
The peak amplitude of the helicity-switched $E_{g}$ mode is larger for the thinner flakes.
In addition, we find Raman peaks near the energy levels of infrared (IR)-active $E_u$ and $A_{2u}$ modes, only for monolayer and a few-layers thick flakes.
Interestingly, these peaks manifest themselves only for helicity-switched Raman scattering; they are nearly absent for helicity-conserved scattering.
\end{abstract}

\pacs{}

\maketitle
Conservation of angular momentum plays a vital role in modern spintronic research.
Spin transfer torque results from the transfer of spin angular momentum from conduction electrons to localized electrons that are responsible of magnetism\cite{slonczewski1996jmmm,berger1996prb,ralph2008jmmm}.
Thermally excited spin waves, often referred to as thermal magnons, can carry spin angular momentum\cite{uchida2010nmat,cornelissen2015nphys}. 
Recent studies have shown that such magnons can also be used to control the magnetization direction of magnetic materials\cite{han2019science,wang2019science}. 
Studies have shown that lattice vibrations, i.e. phonons, also possess angular momentum\cite{zhang2014prl,zhu2018science,ishito2023nphys}.
Identifying phonon modes with non-zero angular momentum facilitates understanding on the interaction of phonons with electrons and possibly with magnetism.

Raman spectroscopy is a powerful tool to analyze the structural properties of materials such as chemical bonds, lattice dynamics, and defects\cite{ferrari2013nnano}.
It has been widely employed in the study of TMDs to identify the crystal symmetry and the number of layers\cite{ferrari2006prl,graf2007nanolett,berkdemir2013scirep}. 
One may select the helicity of the incoming and scattered lights in the Raman spectroscopy measurements.
Such approach has been used to study the nature of phonon(s) involved in the excitation process\cite{chen2015nanolett,tastumi2018prb,zhao2020acsnano}.
For example, the existence of chiral phonons\cite{ishito2023nphys} have been identified using helicity resolved Raman spectroscopy.
Studies have shown that even the magnetic state of magnetic materials be explored using such technique\cite{tian20162dmater,milosavljevic2019prb,lyu2020nanolett}.

PtSe$_2$\cite{wang2015nanolett,cao2021nanotechnol}, air-stable van der Waals material, is reported to possess unique layer-dependent electronic states. 
It has a type-II Dirac band in its bulk phase\cite{sun2020commphys} and becomes an indirect band gap semiconductor when the film is thinned down to a few-layer thickness\cite{huang2016prb,li2017prm,zhang2017prb,yao2017ncomm}
Most notably, recent studies have shown that PtSe2 exhibits layer-dependent magnetic orderings in a few layers-thick state\cite{avsar2020ncomm,manchanda2021prb}.
(The magnetic ordering was attributed to Pt defects\cite{avsar2020ncomm,manchanda2021prb}.)
PtSe$_2$ therefore provides a playground to study the interaction between electrons, phonons and magnetism.

Here we use Raman spectroscopy to study the phonon characteristics of PtSe$_2$ crystals.
We employ helicity resolved Raman spectroscopy to characterize the angular momentum of the phonons that can be excited in monolayer, a few layers-thick and bulk-like PtSe$_2$.
The results are analyzed using the conservation law of angular momentum.
\begin{figure}[b]
\centering
  \includegraphics[width=1\columnwidth]{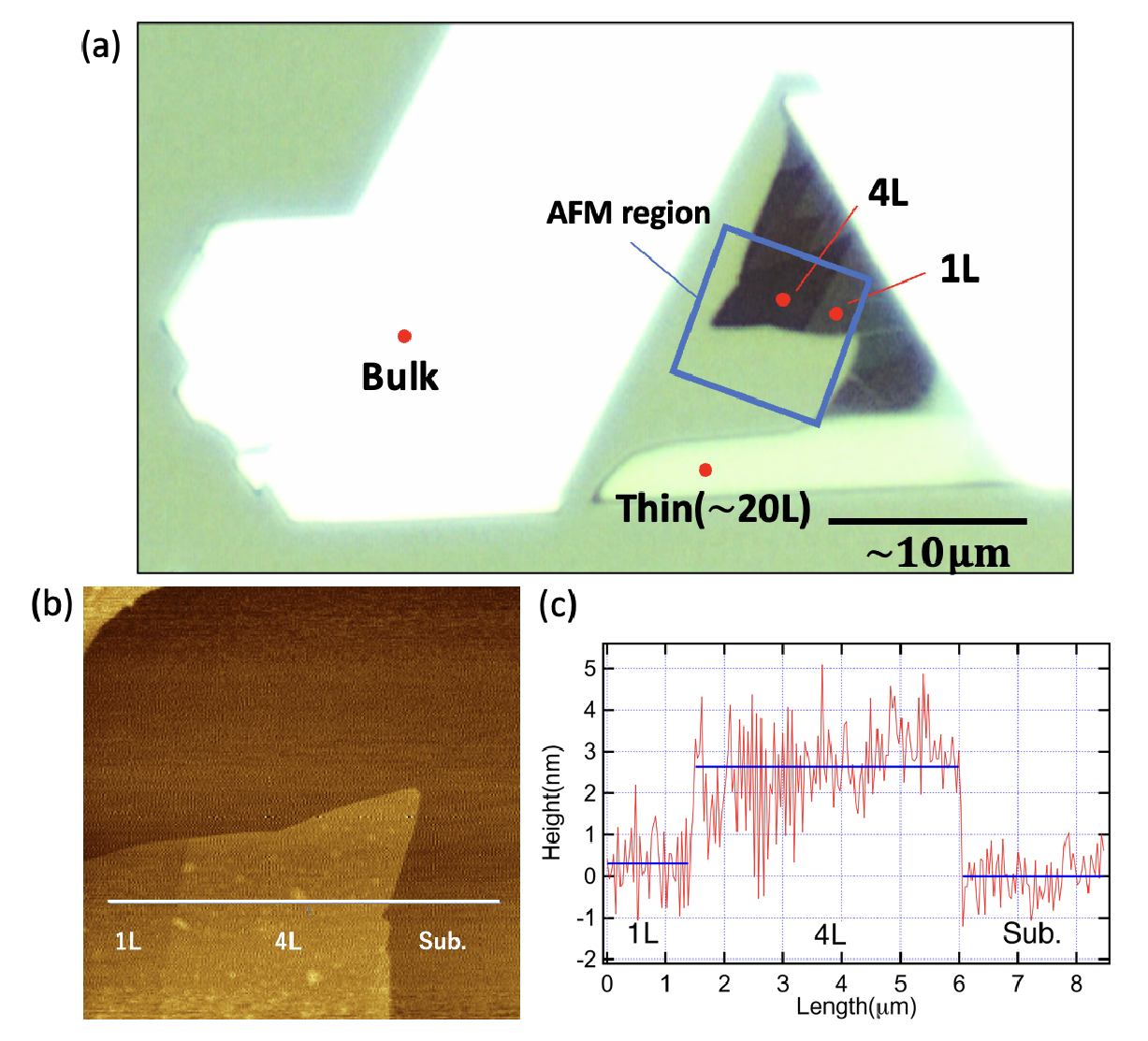}
  \caption{(a) Optical microscope image of the PtSe$_2$ flake used in this study. (b) Atomic force microscopy image of the square region shown in (a). (c) Line profile along the solid thin line in (b).}
  \label{fig:afm}
\end{figure}

PtSe$_2$ flakes are mechanically exfoliated, using adhesive tapes, from commercially available bulk crystals to a polydimethylsiloxane (PDMS) sheet.
The flakes on the PDMS sheet are released on highly doped Si-substrates (covered with 285 nm-thick thermally oxidized silicon) using a conventional transfer process.
Figure~\ref{fig:afm}(a) shows an exemplary optical microscopy image of a flake placed on the Si-substrate.
The bright and dark contrasts represent the reflectivity of the flake that depend on its thickness. 
We use atomic force microscopy to determine the thickness.
The surface topography of the area indicated by the blue square in Fig.~\ref{fig:afm}(a) is shown in Fig.~\ref{fig:afm}(b).
(A rotated view of Fig.~\ref{fig:afm}(a) is shown in Fig.~\ref{fig:afm}(b).)
The image contrast represents the height of the structure.
A line profile along the white solid line in Fig.~\ref{fig:afm}(b) is presented in Fig.~\ref{fig:afm}(c).
From the profile, we determine the thickness of regions of the flake with different contrast. 
The flake thickness of the left-most region is $\sim$0.3 $\pm0.2$ nm, which is obtained from difference in the average height
between the left-most and right-most (substrate) regions.
Taking into account the relatively large error bars of the line profile, such range corresponds to the thickness of a monolayer PtSe$_2$ previously reported ($\sim$0.5 to 0.6 nm)\cite{soled1976matresbull,yan20172dmater}.
The thickness of the middle region is estimated as $\sim$2.6 $\pm0.2$ nm, which roughly corresponds to 4-5 layers.
We denote the thickness of the middle region to be 4 layers hereafter. 
The relation between the film thickness and the image contrast is displayed in Fig.~\ref{fig:afm}(a).

\begin{figure}[b]
\centering
  \includegraphics[width=1\columnwidth]{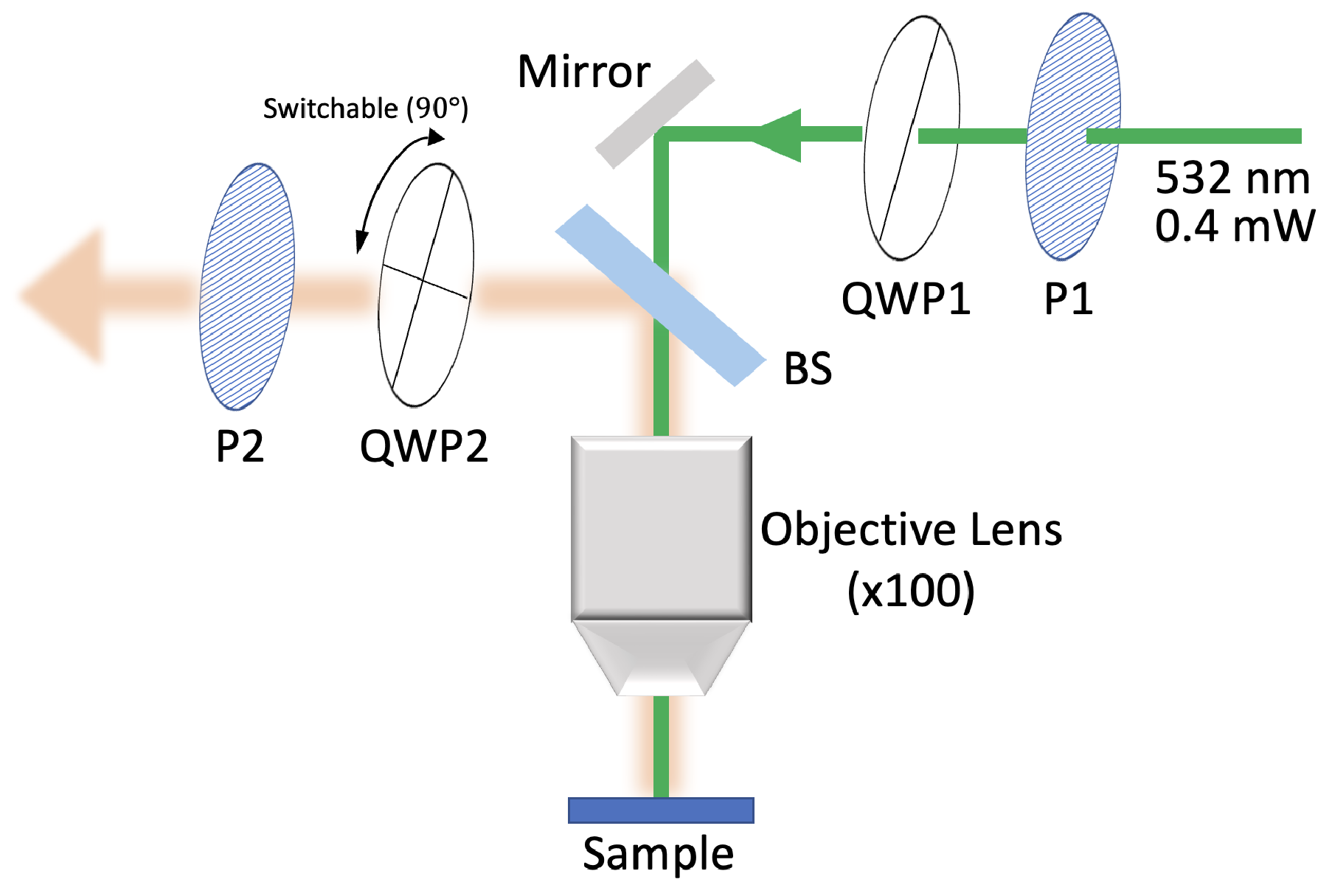}
  \caption{Schematic illustration of the optical setup. BS: beam splitter, P: polarizer, QWP: quarter wave plate.}
  \label{fig:setup}
\end{figure}

Helicity resolved Raman spectroscopy (JASCO: NRS-4500) is performed in the back scattering configuration at room temperature.
The wavelength and the power of the excitation laser are $\sim$532 nm and $\sim$0.4 mW, respectively.
Schematic of the optical setup is illustrated in Fig.~\ref{fig:setup}.
A wire grid polarizer (P1) and a quarter wave plate (QWP1) are inserted between the sample and the excitation laser.
Upon reflection from the sample, the light passes through a second quarter wave plate (QWP2) and a polarizer (P2) and is collected using an aberration-corrected Czerny-Turner type monochromator.
The optical axis of QWP1 and QWP2 is varied to change the helicity of the incident and reflected light.

\begin{figure}[b]
\centering
  \includegraphics[width=1\columnwidth]{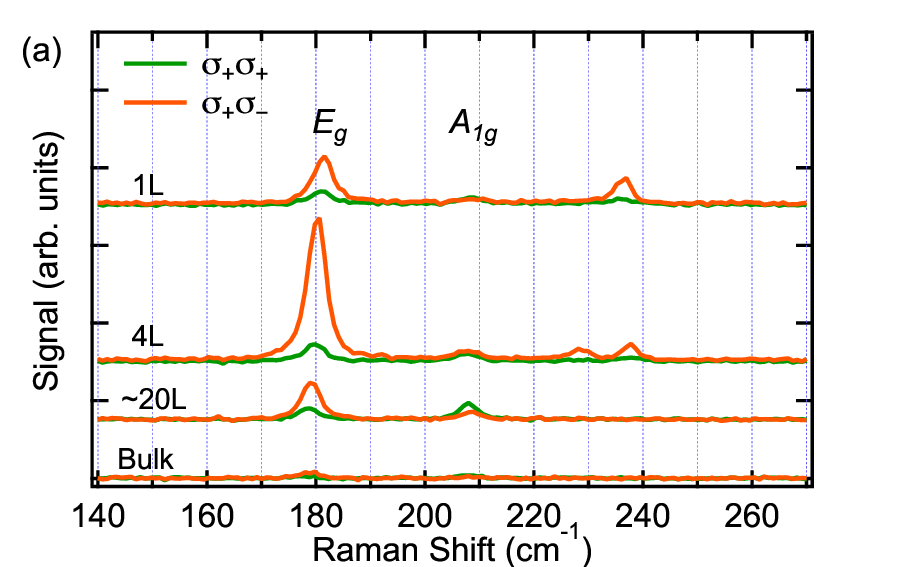}
  \includegraphics[width=1\columnwidth]{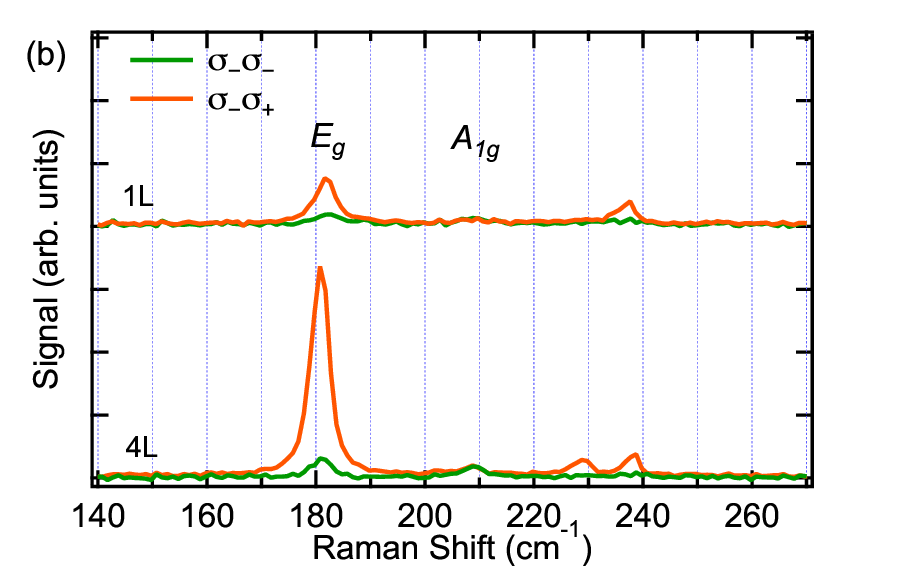}
  \caption{(a,b) Raman spectra of PtSe$_2$ flakes with different thicknesses. The thickness of the flakes are indicated on the left. The green (orange) lines show data when the helicity of the incident and scattered circularly polarized light is the same (opposite). The helicity of the incident light is right-handed (a) and left-handed (b). 
  }
  \label{fig:raman}
\end{figure}

Figures~\ref{fig:raman}(a) and \ref{fig:raman}(b) show the helicity dependent Raman spectra for right- and left-handed circularly polarized incident light, respectively.
The data are vertically shifted for clarity.
The green (orange) solid lines show the Raman spectra that are obtained when the helicity of the incident and reflected lights are the same (opposite).
We refer to such Raman peaks as helicity-conserved and helicity-switched scatterings, respectively.
The characteristics of the spectra are essentially the same for right- and left-handed circularly polarized incident light [Figs.~\ref{fig:raman}(a) and \ref{fig:raman}(b)].
The spectra represent the raw data obtained in the experiments (the laser spot size is kept nearly the same during the measurements of flakes with different thicknesses). 
The amplitude of the peaks is larger for the 4 layers-thick-flake compared to that of the monolayer, which is likely due to the increased volume.
The peak amplitude drops for the ~20 layers-thick-flake, which we infer is caused by incoherent excitation of the vibration modes among the layers.
In the supplementary material, we show the normalized spectra (each spectrum is normalized by the height of the maximum peak) to illustrate the relative amplitude of the peaks.
 
First we focus on the Raman peaks that appear at $\sim$180 \kaiser and $\sim$208 \kaiser.
In accordance with previous studies, these modes correspond to the in-plane $E_g$ and out-of-plane $A_{1g}$ vibrational modes at the center of the Brillouin zone, respectively\cite{obrien20162dmater}.
(From first principles calculations, the $E_g$ and $A_{1g}$ modes can be excited at $\sim$169 cm$^{-1}$ and $\sim$200 cm$^{-1}$, respectively\cite{kandemir2018sst}.)
The peak amplitude of the $E_g$ mode is significantly larger for the helicity-switched scattering than the helicity-conserved scattering.
In contrast, the peak amplitude of the $A_{1g}$ mode is slightly larger for the helicity-conserved scattering.

These characteristics are consistent with the symmetry of the phonons involved in the scattering process.
The most stable structure of PtSe$_2$ is the 1T phase (ABC stacking, the point group is $D_{3d}$).
For this structure, the Raman tensor $\hat{R}$ of the $E_g$ and $A_{1g}$ modes are expressed as\cite{ribeirosoares2014prb}
\begin{equation}
\begin{aligned}
\label{eq:raman:tensor}
\hat{R} (A_{1g})=
\begin{bmatrix}
a & 0 & 0\\
0 & a & 0\\
0 & 0 & b
\end{bmatrix}, \ \ 
\hat{R} (E_{g})=
\begin{bmatrix}
c & 0 & 0\\
0 & -c & d\\
0 & d & 0
\end{bmatrix},
\end{aligned}
\end{equation}
where $a, b, c, d$ are material dependent constants.
The Raman scattering amplitude of an incident state $\ket{\sigma_i}$ to a scattered state $\ket{\sigma_s}$ is given by $\bra{\sigma_s} \hat{R} \ket{\sigma_i}$.
For helicity-conserved and helicity-switched scatterings, the Raman scattering amplitudes are expressed as $\bra{\sigma_\pm} \hat{R} \ket{\sigma_\pm}$ and $\bra{\sigma_\mp} \hat{R} \ket{\sigma_\pm}$, respectively.
Here, $\ket{\sigma_\pm}$ represents the helicity of a circularly polarized light: $\ket{\sigma_\pm} = \frac{1}{\sqrt{2}} (1, \pm i, 0)$.
Using the expression shown in Eq.~(\ref{eq:raman:tensor}), we obtain $\bra{\sigma_\pm} \hat{R}(A_{1g}) \ket{\sigma_\pm} = a$ and $\bra{\sigma_\mp} \hat{R}(A_{1g}) \ket{\sigma_\pm} = 0$ for the $A_{1g}$ mode and $\bra{\sigma_\pm} \hat{R}(E_{g}) \ket{\sigma_\pm} = 0$ and $\bra{\sigma_\mp} \hat{R}(E_{g}) \ket{\sigma_\pm} = c$ for the $E_{g}$ mode. 
Thus the $A_{1g}$ out-of-plane vibration mode is defined by helicity-conserved scattering whereas the $E_g$ in-plane vibration mode is dominated by helicity-switched scattering.
These features are consistent with previous studies on TMDs\cite{chen2015nanolett}.
Note that here we do not observe significant photoluminescence peak under the laser used ($\lambda \sim 532$ nm), indicating that the excitation is non-resonant and therefore Fr\"{o}hlich interaction\cite{zhao2020acsnano} is likely absent.
We consider the Raman tensor analyses is justified under such circumstance.

Recent theoretical studies have shown that the helicity dependent Raman spectra can be accounted for using the conservation law of angular momentum.
Let $\sigma_i$ and $\sigma_s$ be the (spin) angular momentum of the incident and scattered lights, respectively ($\sigma_i$, $\sigma_s =\pm1$).
In addition, we denote the phonon angular momentum by $m_v^\mathrm{ph}$.
The conservation law then dictates the following relation\cite{tastumi2018prb}:
\begin{equation}
\begin{aligned}
\label{eq:conservation:angmom}
\sigma_s - \sigma_i = - m_v^\mathrm{ph} + N_v p,
\end{aligned}
\end{equation}
where $N_v$ is rotational symmetry of the vibrational mode and $p$ is an integer.
The in-plane $E_g$ mode has $N_v = 1$ while the out-of-plane $A_{1g}$ mode has $N_v = 3$.
$m_v^\mathrm{ph}$ is in general 0 for vibration modes that are non-degenerate.
For degenerate modes, $m_v^\mathrm{ph}$ can take values of 0, $\pm1$.
Here the in-plane $E_g$ mode is doublly generate while the out-of-plane $A_{1g}$ mode is non-degenerate: thus $m_v^\mathrm{ph} = 0,\pm1$ for $E_g$ and $m_v^\mathrm{ph} = 0$ for $A_{1g}$.
By substituting $N_v$ and $m_v^\mathrm{ph}$ into Eq.~(\ref{eq:conservation:angmom}), we obtain the allowed values of $\sigma_s - \sigma_i$: the results are shown in the column "$\sigma_s - \sigma_i$" in Table~\ref{table:angmom}.
The selection rule for helicity-conserved and helicity-switched scatterings is determined by examining whether any integer $p$ exists that satisfy $\sigma_s - \sigma_i = 0$ and $\sigma_s - \sigma_i = \pm2$, respectively.
The columns "$\sigma_s - \sigma_i = 0$" and "$\sigma_s - \sigma_i = \pm2$" show the selection rule for helicity-conserved and helicity-switched scatterings, respectively.
Here the circles (crosses) indicate the process is allowed (prohibited). 
From the selection rule, we find both helicity-conserved and helicity-switched scatterings are allowed for the in-plane $E_g$ mode whereas only the helicity-conserved scattering is allowed for the out-of-plane $A_{1g}$ mode.
The latter is consistent with the experimental results and the Raman tensor analysis shown above.
(From the selection rule, it is difficult to determine the dominant vibration mode (helicity-switched or helicity-conserved) for the $E_g$ mode.)
Note that modes with non-zero $m_v^\mathrm{ph}$ correspond to phonons with (pseudo) angular momentum.
\begingroup
\setlength{\tabcolsep}{6pt} 
\begin{table}[b]
 \caption{
 Selection rule of the phonon modes in PtSe$_2$.}
 \label{table:angmom}
 \centering
   \vspace{3pt}
  \begin{tabular}{c c c c c c}
   \hline \hline
   mode & $m_v^\mathrm{ph}$ & $N_v$ & $\sigma_s - \sigma_i$ & $\sigma_s - \sigma_i = 0$ & $\sigma_s - \sigma_i = \pm2$\\
   \hline 
   $A_{1g}$ & 0 & 3 & $3p$ & $\circ$ & $\cross$\\
   $E_{g}$ & 0, $\pm 1$ & 1 & $p, p\pm1$ & $\circ$ & $\circ$\\
   $A_{2u}$ & 0 & 3 & $3p$ & $\circ$ & $\cross$\\
   $E_{u}$ & 0, $\pm 1$ & 1 & $p, p\pm1$ & $\circ$ & $\circ$\\
         \hline
  \end{tabular}
\end{table}
\endgroup

We now turn to the the Raman spectral peaks that appear at 226 cm$^{-1}$ and 236 cm$^{-1}$.
The peaks appear only in monolayer and a few layers-thick PtSe$_2$.
With three atoms per unit cell, there are in total nine vibration modes in PtSe$_2$.
Three of the nine modes are acoustic modes and the remaining six are optical modes.
For 1T-PtSe$_2$ monolayer, the irreducible representation of the six optical modes are expressed as\cite{ribeirosoares2014prb}
\begin{equation}
\begin{aligned}
\label{eq:irrep}
\Gamma = E_g + A_{1g} + 2 E_u + 2 A_{2u}.
\end{aligned}
\end{equation}
The $E_u$ and $A_{2u}$ are known as IR-active modes.
According to reported first principles calculations\cite{kandemir2018sst}, the $E_u$ and $A_{2u}$ modes at the zone center preside at energy levels of $\sim$218 cm$^{-1}$ and $\sim$223 cm$^{-1}$, respectively.
Thus the two modes found in experiments at 226 cm$^{-1}$ and 236 cm$^{-1}$ can be associated with such modes.

Note that the crystal structure of PtSe$_2$ has an inversion symmetry regardless of the number of layers (even for a monolayer system, the inversion symmetry holds\cite{li2018prb,kandemir2018sst}).
For such systems, the vibration mode cannot be both Raman and infrared (IR)-active (mutual exclusion rule); that is, one does not find IR-active modes in the Raman spectra and vice versa.
As evident from Figs.~\ref{fig:raman}(a) and \ref{fig:raman}(b), the peaks close to the IR-active $E_u$ and $A_{2u}$ modes are found only for the thin flakes.
If the peaks are indeed associated with the IR-active modes, the results imply that the crystal structure is slightly changed, breaking the inversion symmetry, as the number of layers is reduced.

Interestingly, the amplitude of the Raman peaks at 226 cm$^{-1}$ and 236 cm$^{-1}$ is larger for the helicity-switched scattering.
We assume that the peaks represent the $E_u$ and/or $A_{2u}$ modes and that the same angular momentum conservation law described in Eq.~(\ref{eq:conservation:angmom}) can be applied.
Similar to the $E_g$ and $A_{1g}$ modes, the phonon angular momentum and the rotational symmetry of the $E_u$ and $A_{2u}$ modes are evaluated and listed in Table~\ref{table:angmom}.
From Eq.~(\ref{eq:conservation:angmom}), we find both helicity-conserved and helicity-switched scatterings are allowed for the $E_u$ mode while only the helicity-conserved scatterings are allowed for the $A_{2u}$ mode.
Given that the observed Raman peaks are larger for the helicity-switched scattering, we infer that the observed peaks are associated with the in-plane $E_u$ modes.
We note that high-order Raman modes\cite{basko2009prb,pimenta2007pccp,delcorro2014acsnano,drapcho2017prb} that involve defects and phonons from other high symmetry points in the Brillouin zone can contribute to the Raman peaks at 226 cm$^{-1}$ and 236 cm$^{-1}$.
It is not obvious why such higher order mode, if present, predominantly appears in the helicity-switched scattering.
As the helicity-switched scattering modes indicate excitation of phonons with angular momentum (i.e. $m_v^\mathrm{ph} \neq 0$), thin PtSe$_2$ may serve as a source of mechanical angular momentum that can be excited using circularly polarized light, useful for potential spin-mechanics applications.  
Recent studies\cite{avsar2019nnano} have shown that PtSe$_2$ exhibits magnetic ordering (at low temperature) when its thickness is small. It remains to be seen whether there is correlation between such defect induced magnetism and the appearance of the helicity dependent IR-active modes.

In summary, we have studied helicity dependent Raman scattering in PtSe$_2$ flakes with different thicknesses.
Regardless of the flake thickness, the Raman peak amplitude of the in-plane $E_{g}$ vibration mode is found to be significantly larger for helicity-switched scattering than that of helicity-conserved scattering. 
In contrast, the out-of-plane $A_{1g}$ vibration mode exhibits larger peak amplitude for the helicity-conserved scattering.
Analyses based on the Raman tensor and conservation of angular momentum account for these results.
For monolayer to a few layers-thick PtSe$_2$, we find Raman peaks that possess similar energy with IR-active modes, \textit{i.e.}, the $E_u$ and $A_{2u}$ modes.
Interestingly, the peak amplitude is significantly larger for the helicity-switched scattering than that for the helicity-conserved scattering, which is almost zero.
Assuming that the conservation law of angular momentum can be applied to IR-active modes, we infer that the observed peaks are the in-plane $E_u$ vibration modes.
The results also suggest that phonons with non-zero angular momentum can be excited in monolayer and a few layers-thick PtSe$_2$, which may prove useful for spintronic applications.


\begin{acknowledgments}
This work was partly supported by JST CREST (Grant No. JPMJCR19T3), MEXT Initiative to Establish Next-generation Novel Integrated Circuits Centers (X-NICS) (Grant No. JPJ011438), and JSPS KAKENHI (Grant Nos. 20J21915, 20J20952). H.M. was supported by the JSR Fellowship, the University of Tokyo.
\end{acknowledgments}

\bibliographystyle{apsrev4-1}
\bibliography{ref_031923}



\end{document}